\documentclass[aps,prl,twocolumn,superscriptaddress]{revtex4-1}

\usepackage{graphicx} 
\usepackage{amssymb} 
\usepackage[usenames]{color}
\usepackage{amsmath}
\usepackage{xspace}
\usepackage{longtable}
\usepackage{soul}
\usepackage{epstopdf}
\usepackage{array}
\usepackage{tabularx}
\usepackage{physics}

\usepackage[colorlinks = true,linkcolor = blue,urlcolor  = blue,citecolor = blue,anchorcolor = blue]{hyperref}

\newcommand{\CrNi}{Cr$_7$Ni\xspace}
\newcommand{\CrMn}{Cr$_7$Mn\xspace}
\newcommand{\CuT}{Cu$_3$\xspace}
\newcommand{\Efield}{$E$-field\xspace}
\newcommand{\Efields}{$E$-fields\xspace}
\newcommand{\tE}{$t_\mathrm{E}$\xspace}
\newcommand{\fE}{$\Delta f_\mathrm{E}$\xspace}

\begin{document}

\title{Electric field control of spins in molecular magnets} 
\author{Junjie Liu}
\email{junjie.liu@physics.ox.ac.uk}
\author{Jakub Mrozek}
\affiliation{CAESR, Department of Physics, University of Oxford, The Clarendon Laboratory, Parks Road, Oxford OX1 3PU, UK}
\author{William K.~Myers}
\affiliation{CAESR, Inorganic Chemistry Laboratory, University of Oxford, South Parks Road, Oxford OX1 3QR, UK}
\author{Grigore A.~Timco}
\author{Richard~E.P.~Winpenny}
\affiliation{School of Chemistry and Photon Science Institute, The University of Manchester, Manchester, M13 9PL, UK}
\author{Benjamin Kintzel}
\author{Winfried Plass}
\affiliation{Institut f\"ur Anorganische und Analytische Chemie, Friedrich-Schiller-Universit\"at Jena, Humboldtstra{\ss}e 8, 07743 Jena, Germany}
\author{Arzhang Ardavan}
\email{arzhang.ardavan@physics.ox.ac.uk}
\affiliation{CAESR, Department of Physics, University of Oxford, The Clarendon Laboratory, Parks Road, Oxford OX1 3PU, UK}

\begin{abstract}
Coherent control of individual molecular spins in nano-devices is a pivotal prerequisite for fulfilling the potential promised by molecular spintronics. By applying electric field pulses during time-resolved electron spin resonance measurements, we measure the sensitivity of the spin in several antiferromagnetic molecular nanomagnets to external electric fields. We find a linear electric field dependence of the spin states in \CrMn, an antiferromagnetic ring with a ground-state spin of $S=1$, and in a frustrated \CuT triangle, both with coefficients of about $2~\mathrm{rad}\, \mathrm{s}^{-1} / \mathrm{V} \mathrm{m}^{-1}$. Conversely, the antiferromagnetic ring \CrNi, isomorphic with \CrMn but with $S=1/2$, does not exhibit a detectable effect. We propose that the spin-electric field coupling may be used for selectively controlling individual molecules embedded in nanodevices.
\end{abstract}
\maketitle

Among the physical manifestations of electronic quantum spins, molecular systems exhibit a range of advantages for technological applications: molecular spin properties may be tailored chemically for particular purposes whilst retaining substantial quantum coherence times~\cite{Ardavan2007,Wedge2012,Zadrozny2015,Atzori2016,Shiddiq2016}; supra-molecular chemical synthesis offers routes to complex multi-spin structures~\cite{REPW_NComms2006,Bowen2015}; and the principles by which individual molecular spin states may be interrogated have been demonstrated~\cite{Burzur2012,Thiele2014,Misiorny2015,Moreno-Pineda2018}. These achievements represent significant progress towards realizing the promise of molecular quantum spintronics~\cite{Leuenberger2001,Troiani2005,Lehmann2007,Bogani2008}. 

The canonical approach to manipulating molecular spins exploits the Zeeman interaction between an oscillatory magnetic field and electron spins~\cite{SchweigerJeschke}. However, this approach to controlling individual molecular components in an integrated device is challenging because it is difficult to localize magnetic fields with the spatial resolution required (on the scale of 1~nm), and undesirable cross-talk is inevitable. On the other hand, electrically-controllable molecular spins~\cite{Kane1998,Trif2008,Trif2010,Islam2010} would offer significant architectural advantages: strong electric fields (\Efields) can be generated and localized over small length scales; and it has been shown that projective spin measurements may be achieved using the same electrodes as for coherent spin manipulation~\cite{Pla2012,Thiele2014,Godfrin2017a}. However, the challenge of quantifying \Efield coupling to spins in molecular nanomagnets persists. Unlike charges, localised electron spins couple only weakly to \Efields because spin-orbital interactions scale with the size of orbitals $L$ as $L^3$~\cite{Golovach2006}. This motivates proposals to exploit \Efield effects in antiferromagnetically-coupled molecular nanomagnets, in which the \Efield-induced modifications to exchange interactions can lead to enhanced spin-electric couplings~\cite{Trif2008,Trif2010}.

Study of the effect of \Efields applied to spins in ESR experiments has, in the past, been used as a means of investigating the symmetry of transition metal centers in insulators or organic materials (for example, proteins and enzymes)~\cite{mims1976linear}; a first-order dependence of the spin energy levels on the applied \Efield is indicative of an inversion-symmetry-broken environment. These techniques were later applied to manipulate coherently potential spin qubits in semiconductors~\cite{George2013} and to study \Efield-induced decoherence~\cite{Bradbury2007,Sigillito2015}. More recently, progress has been made in directly applying high-frequency oscillatory \Efields to drive the spins of individual atoms on surfaces~\cite{Baumann2015},
and molecular spins in single-molecule transistors~\cite{Thiele2014,Godfrin2017a}.

\begin{figure*}[t]
\includegraphics[width=2\columnwidth]{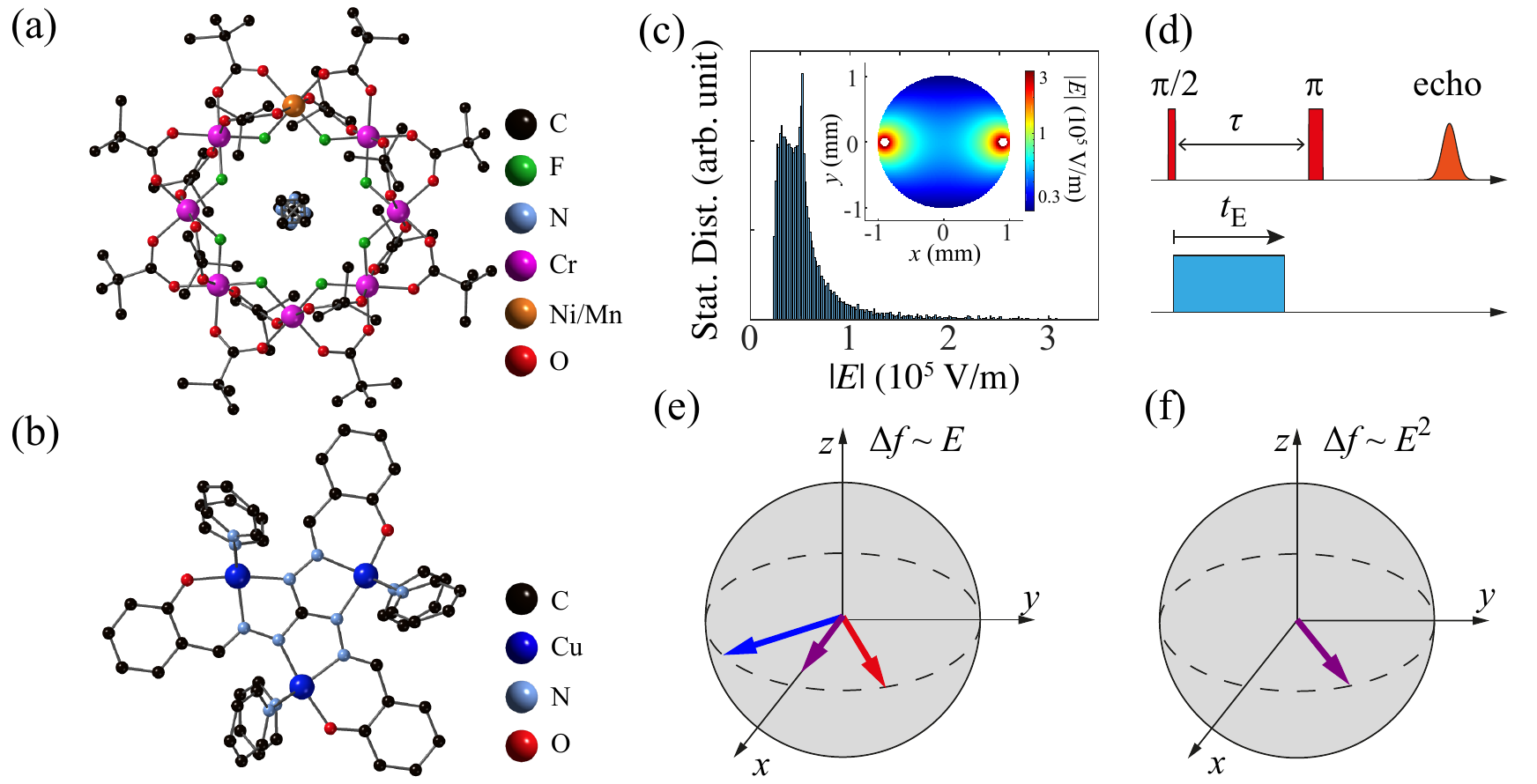}
\caption{(a) The molecular structure of the antiferromagnetic rings \CrNi and \CrMn, which exhibit spin ground states of $S=1/2$ and $S=1$ respectively. (b) The molecular structure of the spin-frustrated \CuT, which exhibits a ground state with $S=1/2$. (c) Calculated distribution for the amplitude of the \Efield, $\vert E \vert$, in the ESR tube with 180 V applied across the electrodes. The colormap illustrates the spatial distribution of $\vert E \vert$ in the cross section of the sample tube. (d) Schematic of the \Efield experimental pulse sequence. A standard Hahn-echo sequence is employed to measure the spin echo signal. An \Efield square pulse of duration \tE is applied immediately after the $\pi/2$ microwave pulse. The echo signal is recorded as a function of the \Efield pulse duration to measure the spin-electric coupling effect. (e) and (f) illustrate the distinguishable effects of linear and quadratic coupling between the \Efield and spins in an orientationally-disordered molecular ensemble. The red and blue arrows in (e) represent the spins of molecules whose orientations are inverted with respect to each other, following a period of free precession under an \Efield; a linear electric field effect gives rise to opposite phase shifts. The echo signal, which results from the sum of all spin packets, remains parallel to the $x$-axis in the rotating frame. For $\Delta f_\mathrm{E} \propto E^2$, the phase shifts for molecules with inverted alignments are identical. Hence, the echo, indicated by the purple arrow in (f), shifts away from the $x$-axis in the rotating frame.}
\label{fig1}
\end{figure*}

In this letter we investigate the spin-electric coupling in a selection of antiferromagnetic molecular nanomagnets, by introducing \Efield pulses to frozen solutions of the molecules during ESR Hahn-echo sequences. Our experiment provides a general method for screening for spin-electric couplings in molecular magnets, thus paving the way for implementation of \Efield-control in molecular spintronics. 

We studied three antiferromagnetic molecules. \CrNi and \CrMn~\cite{Ardavan2007,Wedge2012,Affronte2007,Timco2013}, shown in Figure~\ref{fig1}(a), share the same molecular structure of a ring formed of seven Cr atoms and a heteroatom (either Ni or Mn), bridged by carboxylate ligands~\cite{Cr7M_synthesis}. The antiferromagnetic coupling leads to well-defined magnetic ground states below about 10~K, with a total spin of $S=1$ for \CrMn and $S=1/2$ for \CrNi. The peripheral pivalate groups are deuterated and the molecules are dissolved in deuterated toluene, in order to extend the spin coherence at low temperatures. The \CuT molecule~\cite{Kintzel2018}, shown in Figure~\ref{fig1}(b), is similar to that reported in Reference~\cite{Spielberg2015}, with the bipyridine ligands replaced by pyridine. The molecules are dissolved in deuterated pyridine, which exchanges rapidly with the molecular pyridines, extending the spin coherence at low temperatures. The \CuT core exhibits strong antiferromagnetic interactions between the Cu(II) ions, leading to a $S = 1/2$ ground state at temperatures below 50~K. The structures of all three molecules break inversion symmetry, so they can each, in principle, show a first order \Efield effect~\cite{Mims1974}.

We made ESR measurements using a commercial Bruker Elexsys 580 X-band pulsed ESR spectrometer, equipped with a $^4$He flow cryostat for temperature control. The dissolved samples are contained in standard 3~mm diameter quartz ESR tubes equipped with a pair of electrode wires separated by about 1.8~mm and oriented parallel to the microwave magnetic field, in order to minimize the perturbation to the resonator. To aid impedance matching to the Avtech AVR-4-B voltage pulse generator, the electrodes are shorted above the microwave resonator by a 50~$\Omega$ load, permitting square voltage pulses of up to 180~V with approximately 15~ns rise and fall times, durations up to 30~$\mu$s in 200~ns steps, and a duty cycle of 0.5~\%. This electrode geometry, immersed in the sample solution, generates an inhomogeneous \Efield mostly perpendicular to the microwave magnetic field.  
The distribution of \Efield strengths is shown in Figure~\ref{fig1}(c); when 180~V is applied to the electrodes, most of the sample experiences a field of between $2.5\times10^4$ and $6\times10^4$~V/m. 

The pulse sequence design, shown in Figure~\ref{fig1}(d), is similar to that developed by Mims~\cite{Mims1974}. An \Efield pulse is applied to the sample immediately after the $\pi/2$ pulse in a standard Hahn-echo sequence. If a particular spin packet interacts with the \Efield such that the ESR transition frequency is shifted by \fE, the spins accumulate an extra phase during free precession of $\Delta\varphi_\mathrm{E} = 2\pi \Delta f_\mathrm{E}t_\mathrm{E}$. The inhomogeneity in the \Efield leads to a distribution of phases accumulated across the sample.

\begin{figure*}[t]
\centering
\includegraphics[width=2\columnwidth]{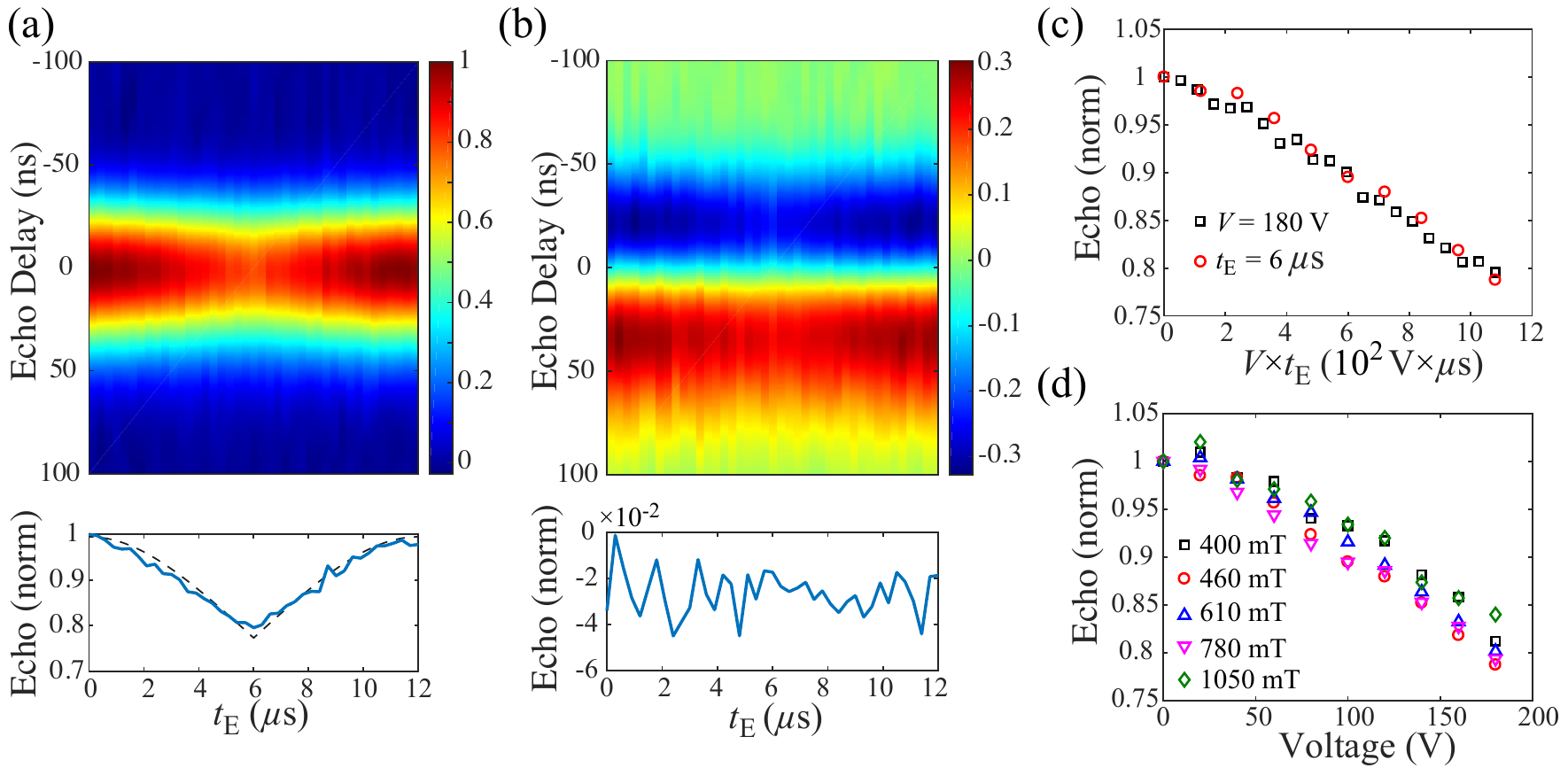}
\caption{\Efield effect on spin echoes for \CrMn. (a) and (b) respectively show the transient in-phase and out-of-phase echoes (colormaps in upper panels) and the integrated echo intensities (lower panels) as a function of the duration of the \Efield pulse (generated by applying a voltage of 180~V to electrodes immersed in the frozen solution).  
(c)~Spin echo intensity as a function of the \Efield pulse. The \Efield pulse (abscissa) is defined as the applied voltage multiplied by its duration. The black squares correspond to measurements with fixed applied voltage (180~V) and varying \Efield pulse duration \tE. The red circles are recorded for a fixed $t_\mathrm{E} = 6~\mu$s while varying the \Efield amplitude. (d)~Dependence of the in-phase integrated echo on \Efield measured at a range of magnetic field positions across the \CrMn ESR absorption spectrum. The temperature is 3~K throughout.}
\label{fig2}
\end{figure*}

There is a further important contribution to the inhomogeneity of the response across the sample, arising from the fact that the molecules' orientations are distributed randomly in the frozen solution. In general, the sensitivity of the molecular spin Hamiltonian depends on the orientation of the molecule with respect to the \Efield. The frequency shift induced on the spin of a molecule with orientation $\mathbf{n}$ is 
$\Delta f_\mathrm{E} = 2\pi \mathbf{n}\cdot\tensor{A}\cdot\mathbf{E}$ 
where $\tensor{A}$ is a second order tensor describing the spin-electric coupling.

Inverting the direction of the \Efield reverses the sign of the phase accumulated as a result of the \Efield pulse. Thus for every spin packet gaining a phase $\Delta\varphi_\mathrm{E}$ in response to the \Efield pulse, there is another spin packet that is shifted by a phase $-\Delta\varphi_\mathrm{E}$, as shown in Figure~\ref{fig1}(e). 

Our experiment measures the ensemble response of the sample, integrated over all molecular orientations excited by the Hahn echo pulses and over the \Efield distribution. The overall effect, for \tE$< \tau$, is to reduce the echo amplitude as a function of the \Efield pulse duration and amplitude, without inducing an out-of-phase component to the echo (Figure~\ref{fig1}(e)). The fact that we measure an averaged response means that we cannot extract from this experiment the detailed structure of $\tensor{A}$. (This would require a sample with orientational order, i.e.\ a crystal; usually, though, dipolar couplings in crystals destroy the phase coherence required for the Hahn echo~\cite{Ardavan2007}.) Instead, we characterise the response of each of the samples studied by an ``average'' isotropic response to the \Efield, and this is the sensitivity figure that we quote.

A second order coupling, i.e.\ \fE $\sim E^2$, leads to the same sign of \fE across the ensemble (see Figure~\ref{fig1}(f)) and therefore a shift of the phase of the spin echo and an out-of-phase component to the echo. We can also check whether the coupling is linear or quadratic by studying the effects of \Efield pulses of different amplitudes; if the coupling is linear, the response should depend only to the product $E$\tE.

Figure~\ref{fig2} shows typical data obtained from \CrMn. Figure~\ref{fig2}(a) and (b) show, respectively, the in-phase and out-of-phase components of the echoes as a function of the duration of the \Efield pulse. The full transients are shown as color-maps in the upper panels and the integrated components of the echoes are plotted in the lower panels. The echo signal is centered at zero delay time, with each vertical cut in the colormaps representing an echo transient for \tE given on the horizontal axis. There is a pronounced modulation of both echo components upon application of the \Efield pulse visible in the transients. The integrated in-phase echo shows a monotonic decrease as \tE, the duration of the \Efield pulse, increases from zero towards $\tau = 6 \mu\mathrm{s}$, where the echo reaches $0.79\pm0.02$ of its \tE$=0$ value. The echo subsequently recovers as \tE exceeds $\tau$ and approaches $2\tau = 12 \mu\mathrm{s}$, because the phase induced by the \Efield during the first period of free evolution is progressively refocussed following the Hahn echo $\pi$-pulse. This confirms that the effect of the \Efield on the molecular spin is coherent. The integrated out-of-phase echo is insignificant throughout. These features are indicative of a linear coupling between the \CrMn spin and the applied \Efield. Figure~\ref{fig2}(c) shows that the echo intensity depends only on the product $V$\tE, confirming that the spin-electric coupling is linear.

The dashed line in the lower panel of Figure~\ref{fig2}(a) is a fit of the echo intensity given the \Efield distribution shown in Figure~\ref{fig1}(c) assuming an isotropic \Efield sensitivity, $A$, yielding $A = 1.9 \pm 0.1~\mathrm{rad}\, \mathrm{s}^{-1} / \mathrm{V} \mathrm{m}^{-1}$. This corresponds to an average phase shift across the ensemble of $\Delta \varphi = \arccos(0.79) \approx 0.66~\mathrm{rad}$, implying an averaged $\Delta f = \Delta \varphi/(2\pi\tau) = 17~\mathrm{kHz}$, or about $1.8\times 10^{-6}$ of the ESR microwave frequency (9.5~GHz).

The $S=1$ \CrMn spin exhibits a substantial anisotropy (with an axial term of about 20~GHz and a rhombic term about an order of magnitude smaller~\cite{Ardavan2007}) such that at 9.5~GHz, the ESR spectrum of the disoriented ensemble extends from about 0.35~T to a little over 1~T. Figure~\ref{fig2}(d) shows the dependence of the integrated in-phase echo on voltage for a fixed \tE measured at several points in the spectrum (corresponding to orientational sub-populations). When normalized to the amplitude of the echo in the absence of an \Efield pulse, we find that there is only a weak dependence on magnetic field. In principle, this type of measurement allows us to separate the dependence of different spin-Hamiltonian terms on \Efield. However, the field dependence that we measure in our disoriented ensemble is not sufficiently distinctive to do this reliably; measurements on oriented ensembles would provide stronger data.

\begin{figure}[t]
\centering
\includegraphics[width=\columnwidth]{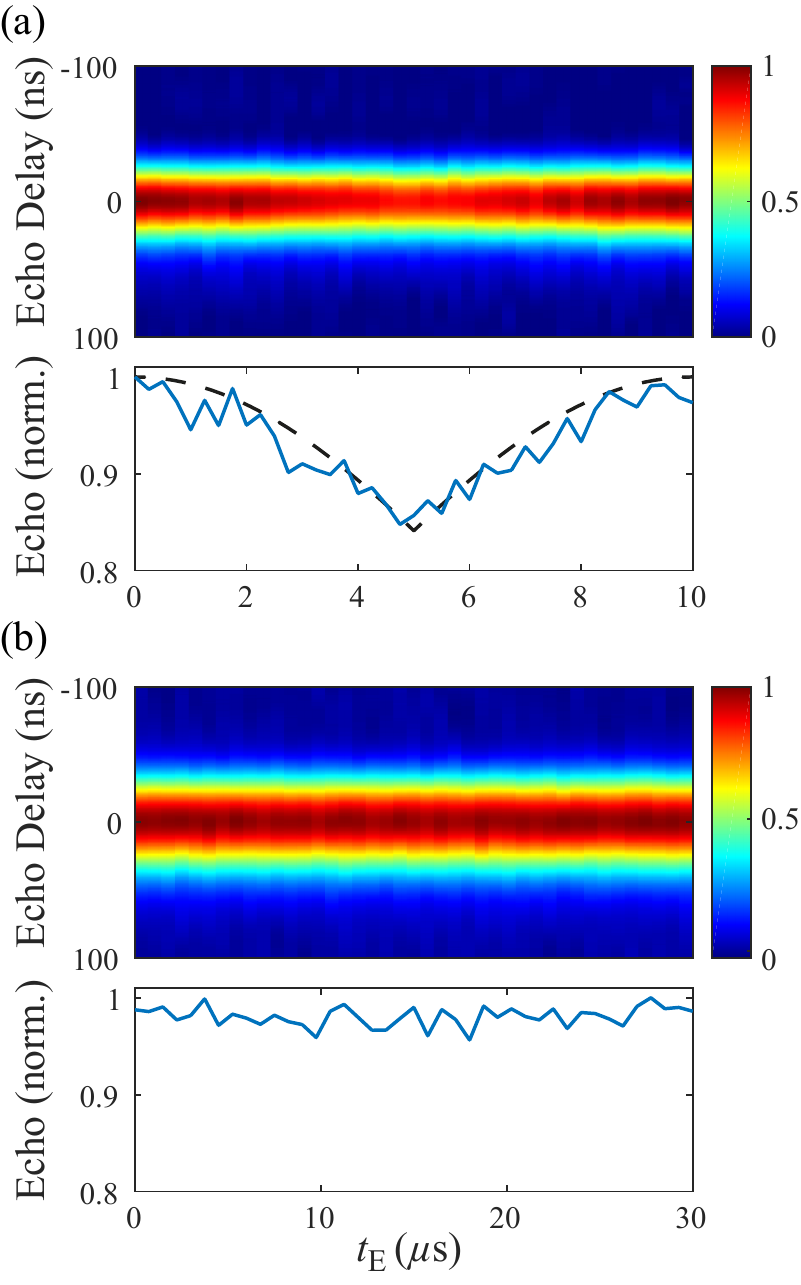}
\caption{The in-phase transient (upper) and integrated (lower) echoes for (a) \CuT and (b) \CrNi as a function of the duration of the \Efield. The amplitude of the \Efield pulses is fixed at 180~V in both cases.  $\tau = 5~\mu$s and 15~$\mu$s in the microwave Hahn-echo sequences for \CuT and \CrNi, respectively. The temperature is 3~K.}
\label{fig3}
\end{figure}

Experiments on \CuT revealed a comparable dependence on \Efield (see Figure~\ref{fig3}(a)), with the normalised echo decreasing to $0.85\pm0.03$ of its \tE$=0$ value for an \Efield pulse of \tE$=5~\mu{\mathrm s}$. Fitting the echo amplitude (as above, dashed line in the lower panel of Figure~\ref{fig3}(a)) yields a spin-electric coupling of $A=1.9 \pm 0.2~\mathrm{rad}\, \mathrm{s}^{-1} / \mathrm{V} \mathrm{m}^{-1}$. 

The data from \CrNi (Figure~\ref{fig3}(b)) show no evidence of spin-electric coupling. The lack of an effect in \CrNi, which shares its broken-inversion-symmetry structure with \CrMn, is interesting. The key difference between \CrNi and \CrMn is the total spin in the ground state ($S=1/2$ and $S=1$ respectively). This suggests that the axial and rhombic zero field splitting anisotropy terms, relevant for \CrMn but not for \CrNi, may be important in offering a sensitivity to the \Efield, as was found, for example, in Mn defects in ZnO~\cite{George2013}. (We note here that the other contribution to the spin anisotropy, through the $g$-factor, is rather weak in both \CrNi and \CrMn.) In this picture, the \Efield sensitivity exhibited by \CuT might be associated with the magnetic frustration inherent in its structure, as proposed by Trif {\it et al}.~\cite{Trif2008,Trif2010}. On the other hand, we note the strength of the \Efield effects in \CrMn and \CuT are comparable to the Stark effect in P donors in Si~\cite{Bradbury2007}, where a quadratic \Efield sensitivity enters through the hyperfine interaction. 
In principle, {\em ab initio} methods can help to explain and predict the magnitude of the \Efield sensitivity~\cite{Trif2010} as shown in Ref.~\cite{Islam2010} for a different \CuT molecular cluster~\cite{Islam_note}. 

The magnitudes of the spin-electric couplings that we observe in \CrMn and \CuT are such that it would be challenging to use them for direct manipulation of the spin via microwave modulation of the \Efield~\cite{George2013,Baumann2015}, but they could be adequate for tuning the ESR transition frequency on short timescales (c.f., the ``\textit{A}-gate'' proposed by Kane for P donors in Si~\cite{Kane1998,Laucht2015}). For example, with sufficiently localized electrodes, the resonance frequency of an individual \CrMn molecule can be shifted by $\sim$32~MHz with an \Efield of $10^8$~V/m, 
an \Efield routinely accessible in reported molecular break-junction devices~\cite{Thiele2014,Burzur2012}. Such control of the resonance frequency is sufficient to shift the molecule in- or out-of-resonance with a globally-applied 30~ns microwave pulse, 100 times shorter than the coherence time of \CrMn molecules. This would fulfil the requirement of selectively controlling individual molecules while achieving rapid spin manipulation using globally-applied microwave pulses.

The spin-electric coupling in molecular nanomagnets might be enhanced through prudent choice of molecular spin centers. For example, rare earth ions with large spin-orbit or hyperfine interactions exhibit sufficient couplings to achieve coherent \Efield-driven manipulations~\cite{Thiele2014,Godfrin2017a}. Alternatively, designing exchange-coupled metal clusters with significant electric dipole moments might offer a route to \Efield tuning of the intra-molecular exchange interactions, and therefore \Efield control over ground-state spin properties. {\em Ab initio} methods can help to guide such rational design~\cite{Islam2010}, for example by identifying intra-molecular exchange interactions that are particularly sensitive to externally-applied \Efields. The experimental method reported here will be important in offering rapid screening of \Efield sensitivities.

We note that a study of the \Efield effect on the continuous-wave ESR spectrum of a crystal of a polynuclear molecular nanomagnet was reported after the submission of this article~\cite{Boudalis2018}.

We thank EPSRC (grants EP/L011972/1, EP/P000479/1, EP/R043701/1), the European Project SUMO (QuantERA call 2017), and European COST Action CA15128 MOLSPIN for support. 

%

\end{document}